\documentclass[lincst,sechang,a4paper]{svmultln}
\usepackage[latin1]{inputenc}
\usepackage{amsmath}
\usepackage{amsfonts}
\usepackage{amssymb}
\usepackage{graphicx}
\usepackage{makeidx}
\makeindex
\begin{document}
\mainmatter

\title{Measuring Accuracy of Automated Parsing and Categorization Tools and Processes in Digital Investigations}
\titlerunning{Measuring Accuracy of Digital Investigations}
\author{}
\institute{}
\author{Joshua I. James\inst{1} \and Alejandra Lopez-Fernandez\inst{2} \and Pavel Gladyhsev\inst{3}}
\authorrunning{Joshua I. James et al.}
\toctitle{Measuring Accuracy of Automated Parsing and Categorization Tools and Processes in Digital Investigations}
\tocauthor{Joshua I. James, Alejandra Lopez-Fernandez, Pavel Gladyshev}
\institute{Digital Forensic Investigation Research Laboratory\\
SoonChunHyang University\\
Shinchang-myeon, Asan-si, South Korea\\
\and
Computer Science and Informatics\\
University College Dublin\\
Belfield, Dublin 4, IE\\
\and
Digital Forensic Investigation Research Laboratory\\
University College Dublin\\
Belfield, Dublin 4, IE}

\maketitle

\begin{abstract}
This work presents a method for the measurement of the accuracy of evidential artifact extraction and categorization tasks in digital forensic investigations. Instead of focusing on the measurement of accuracy and errors in the functions of digital forensic tools, this work proposes the application of information retrieval measurement techniques that allow the incorporation of errors introduced by tools and analysis processes. This method uses a `gold standard' that is the collection of evidential objects determined by a digital investigator from suspect data with an unknown ground truth. This work proposes that the accuracy of tools and investigation processes can be evaluated compared to the derived gold standard using common precision and recall values. Two example case studies are presented showing the measurement of the accuracy of automated analysis tools as compared to an in-depth analysis by an expert. It is shown that such measurement can allow investigators to determine changes in accuracy of their processes over time, and determine if such a change is caused by their tools or knowledge.

\end{abstract}

\section{Introduction}
In digital forensics, the verification and error rates of forensic processes are a common topic. This is mostly due to the evidence admissibility considerations brought on as a result of Daubert v. Merrell Dow Pharmaceuticals, 509 US 579 \cite{SupremeCourt1993}. ``The Daubert process identifies four general categories that are used as guidelines when assessing a procedure'' \cite{Carrier2002}. These are procedure Testing, Error Rate, Publication and Acceptance.

Tools are commonly tested and organizations such as the National Institute of Standards and Technology (NIST) have created test methodologies for various types of tools which are outlined in their Computer Forensic Tool Testing (CFTT) project \cite{NIST2010}. But beyond testing, error rates for tools are not often calculated \cite{James2011, Lyle2010,Baggili2007}. The argument has been made that a tested tool with a high number of users must have a low error rate because if there was a high rate of error, users would not use the tool \cite{Guidance2009}. So far this argument appears to be widely accepted, however Carrier \cite{Carrier2002} submits that ``At a minimum this may be true, but a more scientific approach should be taken as the field matures''. Furthermore, Lyle \cite{Lyle2010} states that ``[a] general error rate [for digital forensic tools] may not be meaningful'', claiming that an error rate should be defined for each function. Because of this, and the lack of Law Enforcement's (LE) time and resources \cite{Gogolin2010}, verification of a tool rarely passes beyond the testing phase of the Daubert process. The same can also be said for the investigator's overall examination process. Some groups claim that a Standard Operating Procedure (SOP) should dictate the overall examination process \cite{Jones2008, UKCodesOfPractice2011}. Validation of this process is commonly done by peer review, but according to James and Gladyshev \cite{James2011} peer review does not always take place. They found that none of the survey respondents mentioned any form of objective measurement of accuracy for the examination process. Further, there has been little research in the area of overall examination accuracy measurement.

Forensic examinations are a procedure for which performance measurement, specifically the measurement of accuracy, is not being conducted, for reasons such as concerns about the subjectivity, practicality and even abuse of such measures \cite{James2013a}. Error rates are created for procedures, tools and functions to determine their probability of failure, and also as a measure for which other methods can be compared against. ``\dots[E]rror rates in analysis are facts. They should not be feared, but they must be measured'' \cite{Palmer2002}. This work is a brief introduction to the problem of accuracy measurement in subjective areas such as digital forensic analysis, why it is needed, and how it may allow investigators to identify when their tools or training is becoming outdated.

\subsection{Contribution}
Previous work has shown that current digital forensic investigations do not normally attempt to quantify the accuracy of examinations beyond the percentage error of investigation tools \cite{James2011}. This work proposes the application of previously known information retrieval accuracy measurement methods to measure the accuracy of digital investigation tools and processes. This work demonstrates that application of the proposed method allows investigators to determine accuracy and error rates of automated or manual processes over time. Further, the proposed method allows investigators to determine where error is being introduced: either at the artifact detection or categorization level. Finally, accuracy measurements can be used to compare the accuracy of highly automated tools -- such as those used in `intelligent' triage -- against a human-created `gold standard' to determine how effective such next-generation digital investigation tools are.

\section{Related Work}
Many fields attempt to measure the accuracy of their processes. In Crawford v. Commonwealth, 33 Va. App. 431 \cite{SupremeCourt2000} -- in regards to DNA evidence -- the jury was instructed that they ``\dots may consider any evidence offered bearing upon the accuracy and reliability of the procedures employed in the collection and analysis\dots'' and that ``DNA testing is deemed to be a reliable scientific technique\dots''. Although the technique may be reliable ``occasional errors arising from accidental switching and mislabeling of samples or misinterpretation of results have come to light\dots'' \cite{Thompson2002}. Furthermore, the relatively recent ``Phantom of Heilbronn'' incident has led to questions of not just internal, but also the external processes that may ultimately effect evidence \cite{Obasogie2009, Yeoman2009}. While the DNA testing technique itself has been deemed to be reliable, erroneous results are still possible due to human error. Digital examinations are not much different in this regard. While a tool may be able to accurately display data, that data is not evidence until an investigator, or a human, interprets it as such. No amount of tool testing can ensure that a human interprets the meaning of the returned results correctly. The law in a region being measured may be used to attempt to objectively define the correctness of an investigation; however, correctness in an investigation is somewhat vulnerable to the subjective conclusions of the investigator and their biases.

Information Retrieval (IR) is one area where accuracy measurement is paramount. Much work has been done in the area of IR, and IR accuracy measurement techniques have previously been applied to forensic text sting searching \cite{Beebe2007}, document classification \cite{DeVel2004}, and even fragmented document analysis in digital forensics \cite{Li2006}. The focus, however, has been on the accuracy measurement of particular techniques or tools within the digital examination process, and not for the examination process itself.

\section{Objective Measures of Analysis Performance}
At present, the efficacy of digital forensic analysis is, in effect, a function of the duration of an examination and of the evidence it produces. These factors force investigators to increase their use of automated tools, and explore autonomous systems for analysis \cite{Kim2004}. Many automated digital forensic tools focus on inculpatory evidence, such as the presence of images, leaving the search for exculpatory evidence to the investigator. Also, many investigators are not comparing their automated tools to a baseline performance measure, such as other similar tools or the results of a manual investigation, which could lead to missed evidence and incomplete investigations. Tools are also not the only component in a digital forensic analysis. Even if all data is displayed correctly, the investigator must then interpret the data correctly. As such, a system of accuracy measurement capable of considering both tools and analysis is needed.

Two simple but informative metrics used in Information Retrieval systems are precision and recall \cite{Russell2009}. This work submits that precision and recall measures can be applied to tools and categorization (analysis) processes in digital investigations. An overall performance measure relative to both the precision and recall, called an F-measure, may be used as the score for overall accuracy of the process. This measurement can help to identify fluctuations in overall process accuracy over time. Precision and recall may then be specifically analyzed to determine if there are problems with artifact identification or categorization. Such metrics may lead to more focused training, smarter budgeting, better tool or technique selection and ultimately higher-quality investigations.

The use of precision and recall is suggested rather than current percentage error methods normally employed in digital forensic tool testing. Percentage error is commonly used to determine the error of a particular function of a tool. While percentage error could be used to evaluate the overall error of artifact categorization in an investigation process with various tools, there is no clear indication where error is being introduced. By using precision and recall, precision can be thought of as the investigator's (or automated tool's) ability to properly classify a retrieved artifact. Recall can be thought of as the investigator's (or automated tool's) ability to discover and retrieve relevant artifacts. These scores can then be used to calculate overall accuracy, which can allow not only identification of weaknesses over time but also whether problems are arising from classification or recall challenges.

\subsection{Digital Analysis}
Evidence, as defined by Anderson and Twinning \cite{Anderson2005}, is ``any fact considered by the tribunal as data to persuade them to reach a reasoned belief [of a theory]''. Digital forensic analysis attempts to identify evidence that supports a theory, contradicts a theory, as well as evidence of tampering \cite{Carrier2002}. If an investigator focuses only on inculpatory evidence, it is possible that they could miss a piece of evidence that may prove the innocence of the suspect, and vice versa. Current digital forensic tools help an investigator to view objects that may have possible evidential value, but what that value is -- inculpatory, exculpatory, tampering, or nothing -- is determined manually by the investigator. The investigator must take the type of case, context of the object and any other evidence into account. This means that the identification of evidential artifacts strongly relates to the knowledge of the investigator. For example, in a survey, 67\% of investigators claimed only a basic familiarity with the Microsoft Windows Registry \cite{James2010b}. If an investigator has little or no knowledge of the Microsoft Windows Registry, he or she may not consider it as a source of evidence. In this case the accuracy of the tool may not be in question, but instead the accuracy of the process or investigator. By using precision and recall compared to a gold standard, the accuracy of both the tool and investigator can be measured, allowing an investigator to determine where error is being introduced.

\subsection{Precision and Recall}
The area of Computer Science known as information retrieval, among others, uses methods to measure the accuracy of the information that is retrieved. Two commonly used metrics are precision and recall. As defined by Russell and Norvig \cite{Russell2009}, ``precision measures the proportion of documents in the result set that are actually relevant... [and] recall measures the proportion of all the relevant documents in the collection that are in the result set''. Manning, Raghavan et al. \cite{Manning2008} define the calculation of precision and recall mathematically using the following formulas:\\
\\
\vspace{1em}
$Precision =\frac{\#\  relevant\  items\  retrieved}{\#\  retrieved\  items} = P(relevant|retrieved)$\\
\vspace{1em}
OR\\
\vspace{2em}
$Precision =\frac{|\{relevant\  items\}\cap\{retrieved\  items\}|}{|\{retrieved\  items\}|}$\\
\vspace{1em}
$Recall =\frac{\#\  relevant\  items\  retrieved}{\#\  relevant\  items} = P(retrieved|relevant)$\\
\vspace{1em}
OR\\
$Recall =\frac{|\{relevant\  items\}\cap\{retrieved\  items\}|}{|\{relevant\  items\}|}$\\

Consider a search engine, for example. When a user enters a query, given enough time, a document containing exactly what the user was looking for could be returned from the set. But if the search had a high level of precision, then the number of documents returned (recalled) would be low and would take more time. Search engines, however, attempt to return results as quickly as possible. Because of this, precision is reduced and a higher number of relevant, but possibly less exact, documents are returned.

An accuracy measure relative to both the precision and recall, called an F-measure (F), may be used as the score for overall accuracy of the measured query. The equation for calculating the F-measure is defined by Russell and Norvig (2009) as:

\begin{center}
$F = 2 \cdot \frac{precision \cdot recall}{precision + recall}$
\end{center}

\subsection{Accuracy of Analysis}
This work proposes that precision and recall may also be applied to the measurement of digital forensic analysis. For example, a digital examination can be considered similar to a search engine query. Digital investigators are asking a question, and their tools return a set of artifacts that may be more or less relevant to the question posed. These artifacts are normally analyzed by an investigator to further remove irrelevant artifacts. Finally, artifacts are then tested for relevance in court. For comparison, a baseline of correctness, or `gold standard', must be established. The artifacts found (recalled) can be used to calculate the accuracy of the examination as compared to a baseline standard.

In digital forensics, peer reviewed in-depth examination of a suspect's system by an expert is the level of examination that is normally accepted for use in court. Because the ground truth about evidential artifacts is unknown, this level of examination may not accurately identify all potential artifacts; however, it is the most comprehensive examination method possible. In other words, with an unknown ground truth, an investigator cannot know what he or she has missed, if anything. In this work an artifact is defined as information that supports or denies a hypothesis. The results of an examination (a collection evidential artifacts) are evaluated for admissibility by the court, resulting in a possible subset of artifacts accepted as evidence. From this, the `gold standard' investigators normally strive for will be defined as \textit{the resulting set of evidential artifacts returned during a peer-reviewed examination that are accepted as admissible evidence in court}. However, in this work the gold standard will be defined as the returned and categorized artifacts after a peer-reviewed examination. With this definition, the gold standard is set at the level of a peer-reviewed human investigation. Using this standard, the results of an examination from other investigators, tools or processes may be objectively compared. Likewise, autonomous digital forensic analysis systems may also be measured against the gold standard, and compared to other processes.

Accuracy of analysis for a certain process, investigator or autonomous system can also be averaged over time to evaluate trends. For example, as software used for analysis becomes out of date, new evidential data sources may exist that the software cannot analyze. By measuring the performance of the process over time, the accuracy may decrease, signaling either an issue with the software or the investigator's knowledge about the new data sources.
 
Since the accuracy of tools using precision and recall has been discussed in other works, this paper will focus on a method for investigator and analysis phase accuracy calculation.

\subsubsection{Measuring the Investigation Process}
In digital forensic analysis, the ideal investigator performance is a high precision (no false positives), and a high recall (no false negatives); all found as fast as possible. Essentially, requirements for an investigator are similar to the requirements for an analysis tool, as described by Carrier \cite{Carrier2002}. An investigator that is comprehensive, accurate and whose work is deterministic and verifiable could be considered competent. This means that both high precision and high recall -- high accuracy -- is equivalent to high performance. This work does not take the weight of artifacts into account. That is, no one artifact is considered any more important than any other. By calculating the investigation process's precision and recall for an analysis, compared to the results of a peer-reviewed examination (or acceptance in court), the resulting accuracy measure may be calculated.

Consider an example where the results of a particular process discovered 4 inculpatory artifacts, and 3 exculpatory artifacts for a total of 7 artifacts. During a peer-reviewed examination the gold standard discovered 9 inculpatory artifacts and 1 exculpatory artifact. This means that the given process led to the discovery of 5 relevant artifacts, missed 5 artifacts, and identified two artifacts falsely compared to the gold standard. In this case, since the gold standard may not be the ultimate truth, a human investigator would need to evaluate whether the falsely identified artifacts were, in fact, not relevant. In the case that they were actually false, precision (P) for the process is found to be:

\begin{center}
$P=\frac{\#\ relevant\ items\ retrieved}{\#\ items\ retrieved}=\frac{5}{7} = 0.71$
\end{center}
\vspace{1em}
Recall (R) is found to be:
\begin{center}
$R=\frac{\#\ relevant\ items\ retrieved}{\#\ relevant\ items}=\frac{5}{10}= 0.5$
\end{center}
\vspace{1em}
Finally, the F-measure (F) is found to be:
\begin{center}
$F=2\cdot \frac{P\cdot R}{P+R}=2\cdot \frac{0.71 \cdot 0.5}{0.71 + 0.5}= 0.59$
\end{center}
\vspace{1em}

In this case the process's precision is 0.71 or 71\%. However, if the process led to the discovery of only one artifact, and that artifact was of evidential value, then the process's precision would be 100\%. In digital investigations, it may be possible that one piece of evidence is all that is necessary, but in many cases supporting information may need to be provided. This is why recall is important. A high precision with a low recall means that the process is missing evidence. In the current example the recall is 0.5 or 50\%. This means that the process missed half of the possible artifacts. The F-measure is the relative combination of the precision and recall. In this case, the examination process scored 0.59 or 59\%. This is the process's accuracy score for this analysis.

\begin{figure}
\centering
\includegraphics[width=\textwidth]{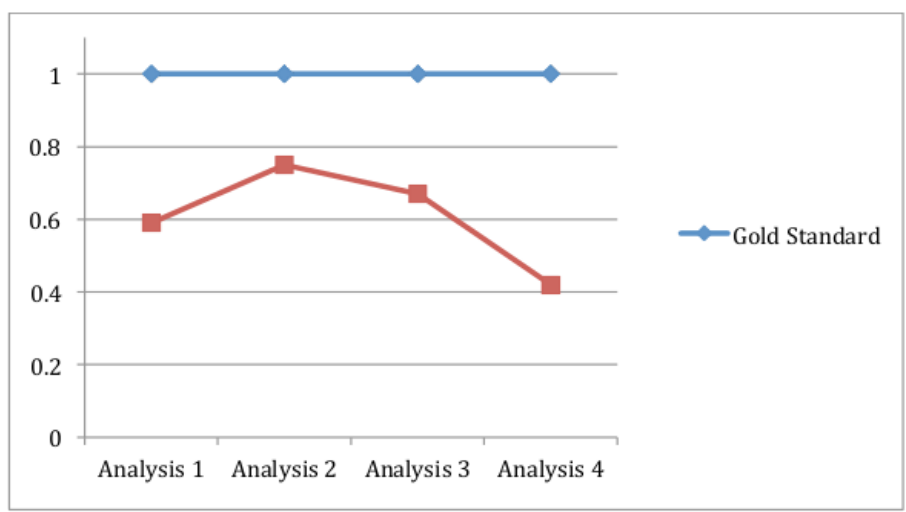}
\caption{Analysis accuracy over time compared to the gold standard}
\label{fig:accuracy_over_time}
\end{figure}

By measuring Precision, Recall and F-measure over time, departments can observe accuracy trends in the examination process, as well as calculate overall performance. Consider the fictional example shown in Table \ref{tab:example_precision}. By examining the F-measure, it can be seen that the process's accuracy is decreasing (Figure \ref{fig:accuracy_over_time}). It can also be seen that the process is consistently missing almost half of the relevant artifacts. By using this method, it becomes easy to see if there are problem areas, and where the problem exists; either with precision (categorization) or recall (artifact extraction).

\begin{table}
\begin{tabular}{|c|c|c|c|}
\hline
& Precision & Recall & F-measure\\ \hline
Analysis 1 & 0.71 & 0.5 & 0.59\\ \hline
Analysis 2 & 1 & 0.6 & 0.75\\ \hline
Analysis 3 & 1 & 0.5 & 0.67\\ \hline
Analysis 4 & 0.7 & 0.3 & 0.42\\ \hline
Average & 0.85 & 0.48 & 0.61\\ \hline
\end{tabular}
\caption{Fictional example calculating Precision, Recall and F-measure for an investigator over time}
\label{tab:example_precision}
\end{table}

\subsubsection{Other Levels of Forensic Examination}
Casey, Ferraro et al. \cite{Casey2002} describe multiple layers of digital forensic examination to help handle an ever-increasing amount of data needing to be analyzed. The use of a multiple layer investigation model has been confirmed by James and Gladyshev \cite{James2011}, where 78\% of respondents claimed to use some sort of preliminary analysis. Most forms of preliminary analysis involve some form of automation, and much of the time if a preliminary analysis is done, the decision to continue or stop the examination will be made based on what is found -- or not -- with these less in-depth processes. It also appears that in all cases if anything suspicious is found during a preliminary examination, then an in-depth analysis will normally take place \cite{James2013b}. Current processes, such as triage, have been shown to help reduce the number of suspect machines needing an in-depth examination; however, triage and highly automated preview examinations are not currently as effective as manual in-depth investigations in every situation \cite{Goss2010, Koopmans2013}. The issue then is that decisions to not continue analysis are being made based on a minimum amount of information. Also, investigators conducting preliminary analyses do not know what is potentially being missed since they are not conducting a full examination.

``To reduce the incidence of incorrect conclusions based on unreliable or inaccurate data it is necessary to quantify uncertainty and correct for it whenever possible'' \cite{Casey2002}. The proposed method to measure accuracy may be applied to all layers of examination. If a highly automated tool, such as a triage solution, is being used to make decisions about a system, an F-measure can be calculated for the solution or process as described and compared to the gold standard. Form this, departments can determine the limitations and benefits of their preliminary analysis techniques and particular tools, resulting in more informed decisions about their overall analysis process.

\section{Implementation of Accuracy Measurement in Digital Forensic Laboratories}
Current tool and process verification methods are challenging to implement in practice for the simple reason that testing is a time-consuming task for laboratories that may already be overburdened. Baggili, Mislan, et al. \cite{Baggili2007} proposed a programmatic approach to error rate calculation but also showed concerns about whether an investigator would record required data in a database. Implementing current tool testing methods usually requires a strict process and definition of scope to potentially only test one out of hundreds of functions of a tool. For tool or process testing to be practical, the measurement process must be minimally disruptive to current digital investigation processes.

While there are many ways to implement the proposed accuracy measurement method in a digital investigation, this work will give one example of how such a measurement process could be implemented in a way that is minimally disruptive to current investigation processes. The proposed measurement method was used during the implementation of a new `Preliminary Analysis Unit' as described by James and Gladyshev \cite{James2013b}.

One major concern with implementing a preliminary analysis phase within a real unit is that, as described, without conducting a full investigation of every piece of media the investigators do not know what may be missed. Since a preliminary analysis process is normally highly automated investigators are limited in the use their own intuition to judge whether an investigation should continue even if nothing was found. As was observed, concerns over missing potential evidence caused the investigators to be more conservative in their judgment for suspect data to receive a full investigations.

The goal of accuracy measurement in this case was to evaluate the accuracy of not only the preliminary analysis tool, but the entire preliminary analysis process. In other words, 1) how accurate was the decision for the suspect data to receive a full analysis and 2) how accurate was the quick preliminary analysis in extracting and identifying evidential data. Precision and recall was used to evaluate the latter question.

Since an entirely new process was being implemented, each preliminary analyst conducted a preliminary analysis on each test case. The output of their analysis was a decision, ``yes'' or ``no'', to conduct a full analysis of the suspect device, and the number of pieces of information that they identified as relevant to the case based on the provided output. Pieces of information would normally be extracted files, but could also be the identification of encryption or information in a log file. For example, the Windows Registry could be considered a single data source, but each key within the data source may provide a single piece of information. When implementing the measurement process, a standard definition must be made about what exactly is being measured and how different data are classified.

Each test case, regardless of the preliminary analyst's decision received a full examination. The examiner was not made aware of the preliminary analyst's decision and results. After each of the suspect test devices received a full investigation, the output of the full investigation was whether the suspect device was relevant, ``yes'' or ``no'', and the number of the pieces of information the investigator identified as relevant.

By comparing the results of the preliminary analysis with the results of the full investigation, the precision, recall and accuracy of the preliminary analysis process could be calculated. By calculating the precision and accuracy of the process, a baseline accuracy was set for the preliminary analysis process. From this it became clear that accuracy was largely analyst-specific. When the output of the tool was given to multiple investigators for analysis, each analyst classified the data -- at least slightly -- differently. And, as expected, the analyst with the most experience was the most accurate.

Testing all cases in a similar manner, however, is not sustainable. Once management was satisfied that the preliminary analysis process was fit for their purposes -- and understood where the process failed -- they opted for measurement on a sample set rather than during each case.

It was decided that each preliminary analyst would conduct a preliminary analysis will full measurement of identified pieces of information on every 10 cases. After, each suspect device would be forwarded for a full investigation regardless of the decision of the preliminary analyst. Each device would receive a full investigation (gold standard) and the results would be compared to the output of the preliminary analysis.

By choosing to conduct measurement on only a sample set, the unit could still receive the benefit examining a fewer number of suspect devices while having continual verification of the process built in.

The proposed accuracy measurement implementation process can be summarized in the following steps:
\begin{enumerate}
\item Identify what is being measured
\item Identify the gold standard
\item Identify how the measured process will be implemented in the current investigation work flow
\item Conduct the measured process
\item Conduct the gold standard process
\item Measure the output of the new implementation against the output of the gold standard
\end{enumerate}

To fist understand the process being implemented, we found it necessary to have a test period where analysis of all suspect devices was measured. Once the process was understood, a sample (ideally a random sample) was chosen and measurement was only conducted for that random sample.

\section{Case Study}
In this section, two cases are given where the proposed accuracy measurement method is used. The first case will use data where an investigator was testing a triage tool against a full human investigation. The second case involves five investigators separately testing a different preliminary analysis tool. Comparisons between the investigators, as well as the tools are then evaluated.

\subsection{Case 1}
The following example case has been adapted from the work of Goss \cite{Goss2010}, where the accuracy of a newly implemented triage process is being compared to a human investigator conducting a full analysis on the given media. In this case the accuracy of automated triage analysis will be compared to the gold standard set by an in-depth manual investigation based on the analysis of a data set with an unknown ground truth. In other words, the automatic classification of objects as having evidential value is being compared to the human gold standard. Five automated triage examinations (5 different media) are given in Appendix \ref{Appx:Goss_Results}, with their precision, recall and F-measure calculated. In this case, the human investigator validated the gold standard. For this reason, only false positives, as compared to the gold standard, with no further validation, are given. Table \ref{tab:case1_summary} gives a summary of the examination accuracy results.

\begin{table}
\begin{tabular}{|c|c|c|c|}
\hline
& Precision & Recall & F-measure\\ \hline
Analysis 1 & 0.67 & 0.33 & 0.44\\ \hline
Analysis 2 & 0.00 & 0.00 & 0.00\\ \hline
Analysis 3 & 1.00 & 1.00 & 1.00\\ \hline
Analysis 4 & 0.07 & 0.53 & 0.12\\ \hline
Analysis 5 & 0.15 & 0.12 & 0.13\\ \hline
Average & 0.38 & 0.40 & 0.34\\ \hline
\end{tabular}
\caption{Summary of examination accuracy results using precision and recall to calculate the overall F-measure}
\label{tab:case1_summary}
\end{table}

From Table \ref{tab:case1_summary}, the accuracy of the triage analysis conducted varies greatly. By observing these fluctuations, their cause may possibly be determined. Analysis 2, for example, had poor results because triage is unable to take the context of the case into consideration, and out of context the results returned by a quick triage examination might be suspicious. Alternatively, analysis 3 was extremely accurate because all discovered evidence was found using a `known-bad' hash database, and only previously known artifacts (artifacts that were in the hash database) were on the suspect system. Overall in this case it can be said that this triage tool, as configured, is good for finding known, or `low hanging', artifacts but it is not as effective as an in-depth examination by the investigator during more complicated investigations.

Using this method, it is shown that the overall precision of the implemented triage solution in this particular case study is 38\%, and that it is missing 60\% of the possible evidential artifacts as compared to the gold standard. The overall accuracy `grade' for the implemented triage analysis is 34\%. From here, this measurement can be used as a baseline for improvement, comparison with other automated tools, or to focus when triage should and shouldn't be used. Also, when using this method it becomes clear in which situations triage is missing evidence. With this knowledge, the triage process could possibly be changed to be more comprehensive.

\subsection{Case 2}
The second case involves five pieces of suspect media that each received a full expert digital forensic analysis, and had reports written as to the findings of all evidential artifacts. Each case was an investigation into possession of suspected child exploitation material. After the suspect media received a full manual analysis by an expert investigator, five preliminary examiners conducted a blind analysis on each piece of media using a locally developed preliminary analysis tool. One preliminary examiner (examiner 1) had experience conducting in-depth digital forensic investigations, while the remaining investigators had no experience with in-depth digital forensic analysis. The goal was to determine if decisions to discard media that did not contain illegal material could accurately be made without a time-consuming full examination. To test this method, the decision error rate was examined as well as the preliminary analysis precision rate using the described method to attempt measure the precision of both the tool and the examiner. The results of each preliminary analysis are given in Appendix \ref{Appx:Case2_Results}.

In the context of this case study, false positives are defined as artifacts identified as suspicious, but are in fact not illegal according to the gold standard. False negatives are defined as artifacts that are illegal that were not identified according to the gold standard. It is important to note that in a preliminary analysis it is acceptable -- and likely -- to have false positives in both the object identification and decision for further analysis. This process, however, must have a false negative rate of 0 for the decision for further analysis, meaning that exhibits with illegal content are always sent for further analysis. This process does not necessarily need a false negative rate of 0 for illegal artifact identification, since full illegal artifact identification is the purpose of the full analysis.

Five test cases were given where the suspect media with unknown ground truth received a full manual analysis by a human investigator, from which a report of findings was created. This report is considered the gold standard for classification of artifacts as illegal or unrelated. All cases were based on charges of possession of child exploitation material. Out of the five suspect media, three (60\%) were found to not contain illegal content. Two exhibits (40\%) were found to contain illegal content, most of which were illegal images.

A preliminary examiner then used an automated tool for object extraction purposes, and manually classified objects as illegal or unrelated. Table \ref{tab:case2_analysis_decision} gives the overall results of the preliminary examiner's further analysis decision and accuracy rates, Table \ref{tab:case2_ave_id_error} shows the average artifact identification error rate per preliminary examiner compared to the gold standard, and Table \ref{tab:case2_ave_accuracy} displays the average accuracy rate based on artifact identification per investigator compared to the gold standard.

\begin{table}
\resizebox{\textwidth}{!} {
\begin{tabular}{|c|c|c|c|c|}
\hline
\multicolumn{5}{|l|}{Media Further Analysis Decision Error Rate}\\ \hline
Examiner & False Positive & False Positive Error & False Negative & False Negative Error\\ \hline
Examiner 5 & 2 & .4 & 0 & 0\\ \hline
Examiner 4 & 2 & .4 & 0 & 0\\ \hline
Examiner 3 & 2 & .4 & 0 & 0\\ \hline
Examiner 1 & 1 & .2 & 0 & 0\\ \hline
Examiner 2 & 2 & .4 & 0 & 0\\ \hline
\end{tabular}
}
\caption{Further analysis decision false positive and false negative error rates per preliminary examiner}
\label{tab:case2_analysis_decision}
\end{table}

\begin{table}
\resizebox{\textwidth}{!} {
\begin{tabular}{|c|c|c|c|}
\hline
\multicolumn{3}{|l|}{Average Object Identification Error Rate}\\ \hline
Examiner & Ave. False Positive Error & Ave. False Negative Error\\ \hline
Examiner 5 & .4 & .26\\ \hline
Examiner 4 & .31 & .13\\ \hline
Examiner 3 & .35 & .02\\ \hline
Examiner 1 & .21 & .24\\ \hline
Examiner 2 & .31 & .09\\ \hline
\end{tabular}
}
\caption{Average artifact identification error rate per preliminary examiner}
\label{tab:case2_ave_id_error}
\end{table}

\begin{table}
\begin{tabular}{|c|c|c|c|}
\hline
\multicolumn{2}{|l|}{Average Accuracy Rate}\\ \hline
Examiner & F-measure\\ \hline
Examiner 5 & .35\\ \hline
Examiner 4 & .57\\ \hline
Examiner 3 & .80\\ \hline
Examiner 1 & .64\\ \hline
Examiner 2 & .55\\ \hline
Unit Ave. & .58\\ \hline
\end{tabular}
\caption{Average accuracy rate based on artifact identification per preliminary examiner}
\label{tab:case2_ave_accuracy}
\end{table}

From the Table \ref{tab:case2_analysis_decision}, it is shown that no preliminary examiner falsely excluded suspect media containing illegal material. This means that all exhibits containing illegal material would have received an in-depth analysis. Also, Table \ref{tab:case2_analysis_decision} shows that the preliminary examiner with more experience -- Examiner 1 -- had a lower false positive rate in the decision making process. This is presumably due to a better ability to categorize and differentiate between illegal and borderline content. From Table \ref{tab:case2_ave_id_error}, it can be seen that false positive rates for artifact identification were relatively high. This was an expected outcome since the preliminary examiners are not capable of definitely categorizing borderline illegal content. A higher false positive rate may also indicate the preliminary examiners being overly cautious. Also from Table \ref{tab:case2_ave_id_error}, the false negative rate for artifact identification is somewhat high. This is also expected since preliminary examiners are not conducting a full analysis. Artifact identification false negatives must be compared with the results in Table \ref{tab:case2_analysis_decision}. When comparing artifact identification to the decision process, missing some of the illegal material did not have an effect on the decision process. This is because if there are suspect artifacts, there are likely multiple sources that are suspicious. However, this correlation should be continuously monitored.

\subsection{Evaluation}
Table \ref{tab:case2_ave_accuracy} is the calculated average accuracy rate based on automatic artifact identification and manual classification. This is a metric that may be used for measurement and comparison in the future to ensure continued quality, where recall correlates to the ability of the tool to return related artifacts, and precision correlates to a preliminary examiner's ability to correctly categorize returned artifacts. If each preliminary examiner dropped in accuracy, it may indicate an issue with tools not extracting the required artifacts, or possibly an issue with the training of the preliminary examiner.

The calculated average accuracy rates may also be used to compare two analysis methods. As an example, consider Table \ref{tab:case1_summary}, where the average accuracy of the Case 1 triage solution compared to the gold standard (full analysis) was .34 (34\%). If this is compared to the average calculated accuracy -- .58 (58\%) -- of the (mostly untrained) preliminary examiners in Case 2, it can be seen that the preliminary examiners in Case 2 are approximately .24 (24\%) more accurate than the Case 1 triage solution for making similar decisions. Other metrics, however, should also be considered, such as the time for processing and analysis. For example, the Case 1 triage solution is meant to run on-scene in approximately 2 hours or less, not including analysis. The preliminary analysis solution in Case 2 is designed to be ran in a laboratory from 24 to 48 hours, depending on the size of the suspect media. Because of this, improved accuracy may be expected, but at the cost of time.

\subsubsection{Limitations}
There are two main limitations to the proposed method, the greatest being the definition of the gold standard. The gold standard, as defined in this paper, requires an expert to verify the findings of a given analysis. While such verification is sometimes performed as a matter of course in digital forensic laboratories, not all organizations can afford to duplicate efforts, even on a random sample. Furthermore, it should be noted that a gold standard is only as good as the experts creating it. If a sub-par examiner is setting the standard, the results of measurement may look very good even for poor examinations.

The second limitation is that the accuracy measurement cannot be used when no illegal artifacts were found in the full analysis. This method is only useful in measuring when some objects -- either inculpatory or exculpatory -- are discovered by the gold standard.

\section{Conclusions}
This paper proposed the application of precision and recall metrics for the measurement of the accuracy of digital forensic analyses. Instead of focusing on the measurement of accuracy and errors in digital forensic tools, this work proposed the use of Information Retrieval concepts to incorporate errors introduced by tools and the overall investigation processes. By creating a gold standard with which to compare to, the accuracy of tools and investigation processes can be evaluated, and trends over time determined. From the calculated accuracy it can be determined whether artifact identification or categorization is leading to lower accuracy. This may allow an investigator to assess whether error may lie in the tools or the training over time. The proposed measurement may be applied to many different layers of the investigation process to attempt to determine the most accurate processes, how those processes change over time, and how the unit should change with new trends.

\bibliographystyle{plain}
\bibliography{accuracy}

\newpage
\section*{Appendix}
\appendix
\section{Case 1: Results of precision of investigation vs. the gold standard}
\label{Appx:Goss_Results}

\textbf{Examination 1:}
\begin{table}
\begin{tabular}{|c|c|c|c|c|}
\hline
& Inculpatory & Exculpatory & False Positive & Total\\ \hline
Gold Standard & 12 & 0 & N/A & 12\\ \hline
Triage Examination & 4 & 0 & 2 & 6\\ \hline
\end{tabular}
\caption{Examination 1 artifacts identified compared to the gold standard}
\end{table}

\begin{center}
$P=\frac{4}{6}=0.67$
$R=\frac{4}{12}=0.33$
$F=2\cdot \frac{0.67 \cdot 0.33}{0.67 + 0.33} = 0.44$
\end{center}

\textbf{Examination 2:}
\begin{table}
\begin{tabular}{|c|c|c|c|c|}
\hline
& Inculpatory & Exculpatory & False Positive & Total\\ \hline
Gold Standard & 0 & 1 & N/A & 1\\ \hline
Triage Examination & 0 & 0 & 5 & 5\\ \hline
\end{tabular}
\caption{Examination 2 artifacts identified compared to the gold standard}
\end{table}

\begin{center}
$P=\frac{0}{5}=0$
$R=\frac{0}{1}=0$
$F=2\cdot \frac{0 \cdot 0}{0 + 0} = 0$
\end{center}

\textbf{Examination 3:}
\begin{table}
\begin{tabular}{|c|c|c|c|c|}
\hline
& Inculpatory & Exculpatory & False Positive & Total\\ \hline
Gold Standard & 200 & 0 & N/A & 200\\ \hline
Triage Examination & 200 & 0 & 0 & 200\\ \hline
\end{tabular}
\caption{Examination 3 artifacts identified compared to the gold standard}
\end{table}

\begin{center}
$P=\frac{200}{200}=1$
$R=\frac{200}{200}=1$
$F=2\cdot \frac{1 \cdot 1}{1 + 1} = 1$
\end{center}

\newpage
\textbf{Examination 4:}
\begin{table}
\begin{tabular}{|c|c|c|c|c|}
\hline
& Inculpatory & Exculpatory & False Positive & Total\\ \hline
Gold Standard & 30 & 0 & N/A & 30\\ \hline
Triage Examination & 16 & 0 & 200 & 216\\ \hline
\end{tabular}
\caption{Examination 4 artifacts identified compared to the gold standard}
\end{table}

\begin{center}
$P=\frac{16}{216}=0.07$
$R=\frac{16}{30}=0.53$
$F=2\cdot \frac{0.07 \cdot 0.53}{0.07 + 0.53} = 0.12$
\end{center}

\textbf{Examination 5:}
\begin{table}
\begin{tabular}{|c|c|c|c|c|}
\hline
& Inculpatory & Exculpatory & False Positive & Total\\ \hline
Gold Standard & 34 & 0 & N/A & 34\\ \hline
Triage Examination & 4 & 0 & 22 & 26\\ \hline
\end{tabular}
\caption{Examination 5 artifacts identified compared to the gold standard}
\end{table}

\begin{center}
$P=\frac{4}{26}=0.15$
$R=\frac{4}{34}=0.12$
$F=2\cdot \frac{0.15 \cdot 0.12}{0.15 + 0.12} = 0.13$
\end{center}

\section{Case 2: Results of precision of investigation vs. the gold standard}
\label{Appx:Case2_Results}

\begin{table}
\begin{tabular}{|c|p{23.5em}|}
\hline
\multicolumn{2}{|l|}{Fully-Examined Case}\\ \hline
Suspect Objects & Notes\\ \hline
0 & No illegal content was detected in a full analysis\\ \hline
\end{tabular}
\caption{Results of a full examination on media number 1}
\end{table}

\begin{table}
\resizebox{\textwidth}{!} {
\begin{tabular}{|c|c|c|p{15em}|}
\hline
Examiner & Further Analysis & Suspect Objects & Notes\\ \hline
Examiner 1 & Yes & 4 & Decision made based on found images suspicious deleted files and searching activity\\ \hline
Examiner 2 & Yes & 6 & Decision made based on found images, cleaner programs, Internet activity and evidence of P2P activity\\ \hline
Examiner 3 & Yes & 4 & Decision made based on found images, movies and Internet search and browser history\\ \hline
Examiner 4 & Yes & 30 & Decision made based on large amount of highly suspicious images and some movie files\\ \hline
Examiner 5 & Yes & 903 & Decision made based on a large amount suspicious images\\ \hline
\end{tabular}
}
\caption{Results of preliminary analysis on media number 1 from five examiners}
\end{table}

\begin{table}
\resizebox{\textwidth}{!} {
\begin{tabular}{|c|c|c|c|c|}
\hline
\multicolumn{5}{|l|}{Object Identification Error Rate}\\ \hline
Examiner False & Positive & False Positive Error & False Negative & False Negative Error\\ \hline
Examiner 1 & 4 & 1 & 0 & 0\\ \hline
Examiner 2 & 6 & 1 & 0 & 0\\ \hline
Examiner 3 & 4 & 1 & 0 & 0\\ \hline
Examiner 4 & 30 & 1 & 0 & 0\\ \hline
Examiner 5 & 903 & 1 & 0 & 0\\ \hline
\end{tabular}
}
\caption{Preliminary analysis object identification error rates for media number 1}
\end{table}

\begin{table}
\begin{tabular}{|c|c|c|c|}
\hline
\multicolumn{4}{|l|}{Accuracy Rate}\\
Examiner & Precision & Recall & F-measure\\ \hline
Examiner 1 & n/a & n/a & n/a\\ \hline
Examiner 2 & n/a & n/a & n/a\\ \hline
Examiner 3 & n/a & n/a & n/a\\ \hline
Examiner 4 & n/a & n/a & n/a\\ \hline
Examiner 5 & n/a & n/a & n/a\\ \hline
\end{tabular}
\caption{Preliminary analysis accuracy rates for media number 1}
\end{table}

\begin{table}
\begin{tabular}{|c|p{23.5em}|}
\hline
\multicolumn{2}{|l|}{Fully-Examined Case}\\ \hline
Suspect Objects & Notes\\ \hline
19 & All illegal objects were images\\ \hline
\end{tabular}
\caption{Results of a full examination on media number 2}
\end{table}

\begin{table}
\resizebox{\textwidth}{!} {
\begin{tabular}{|c|c|c|p{15em}|}
\hline
Examiner & Further Analysis & Suspect Objects & Notes\\ \hline
Examiner 2 & Yes & 44 & Decision made based on suspicious images, cookies and installed cleaner\\ \hline
Examiner 5 & Yes & 9 & Decision made based on suspicious images. Note: more suspicious images available not listed in report.\\ \hline
Examiner 1 & Yes & 6 & Decision made based on suspicious movie, porn chat (cookies), possible disk wiping, undeleted, and nothing in the live set\\ \hline
Examiner 3 & Yes & 75 & Decision made based on suspicious images, undeleted and cookies\\ \hline
Examiner 4 & Yes & 40 & Decision made based on many suspicious undeleted images and trace cleaning software\\ \hline
\end{tabular}
}
\caption{Results of preliminary analysis on media number 2 from five examiners}
\end{table}

\begin{table}
\resizebox{\textwidth}{!} {
\begin{tabular}{|c|c|c|c|c|}
\hline
\multicolumn{5}{|l|}{Object Identification Error Rate}\\ \hline
Examiner False & Positive & False Positive Error & False Negative & False Negative Error\\ \hline
Examiner 2 & 25 & .56 & 0 & 0\\ \hline
Examiner 5 & 0 & 0 & 10 & .47\\ \hline
Examiner 1 & 1 & .05 & 13 & .68\\ \hline
Examiner 3 & 56 & .74 & 0 & 0\\ \hline
Examiner 4 & 21 & .53 & 0 & 0\\ \hline
\end{tabular}
}
\caption{Preliminary analysis object identification error rates for media number 2}
\end{table}

\begin{table}
\begin{tabular}{|c|c|c|c|}
\hline
\multicolumn{4}{|l|}{Accuracy Rate}\\
Examiner & Precision & Recall & F-measure\\ \hline
Examiner 2 & .43 & 1 & .60\\ \hline
Examiner 5 & 1 & .47 & .64\\ \hline
Examiner 1 & .83 & .26 & .40\\ \hline
Examiner 3 & .25 & 1 & .41\\ \hline
Examiner 4 & .48 & 1 & .64\\ \hline
\end{tabular}
\caption{Preliminary analysis accuracy rates for media number 2}
\end{table}

\clearpage

\begin{table}
\begin{tabular}{|c|p{23.5em}|}
\hline
\multicolumn{2}{|l|}{Fully-Examined Case}\\ \hline
Suspect Objects & Notes\\ \hline
0 & No evidence or trace evidence relevant to the investigation\\ \hline
\end{tabular}
\caption{Results of a full examination on media number 3}
\end{table}

\begin{table}
\resizebox{\textwidth}{!} {
\begin{tabular}{|c|c|c|p{15em}|}
\hline
Examiner & Further Analysis & Suspect Objects & Notes\\ \hline
Examiner 3 & Yes & 0 & Decision made based on presence of virtual machines\\ \hline
Examiner 5 & No & 0 & n/a\\ \hline
Examiner 4 & Yes & 0 & Decision made based on evidence that user is highly computer literate\\ \hline
Examiner 2 & Yes & 0 & Decision made based on deleted files that could not be processed -- user also highly computer literate\\ \hline
Examiner 1 & No & 0 & n/a\\ \hline
\end{tabular}
}
\caption{Results of preliminary analysis on media number 3 from five examiners}
\end{table}

\begin{table}
\resizebox{\textwidth}{!} {
\begin{tabular}{|c|c|c|c|c|}
\hline
\multicolumn{5}{|l|}{Object Identification Error Rate}\\ \hline
Examiner False & Positive & False Positive Error & False Negative & False Negative Error\\ \hline
Examiner 3 & 0 & 0 & 0 & 0\\ \hline
Examiner 5 & 0 & 0 & 0 & 0\\ \hline
Examiner 4 & 0 & 0 & 0 & 0\\ \hline
Examiner 2 & 0 & 0 & 0 & 0\\ \hline
Examiner 1 & 0 & 0 & 0 & 0\\ \hline
\end{tabular}
}
\caption{Preliminary analysis object identification error rates for media number 3}
\end{table}

\begin{table}
\begin{tabular}{|c|c|c|c|}
\hline
\multicolumn{4}{|l|}{Accuracy Rate}\\
Examiner & Precision & Recall & F-measure\\ \hline
Examiner 3 & n/a & n/a & n/a\\ \hline
Examiner 5 & n/a & n/a & n/a\\ \hline
Examiner 4 & n/a & n/a & n/a\\ \hline
Examiner 2 & n/a & n/a & n/a\\ \hline
Examiner 1 & n/a & n/a & n/a\\ \hline
\end{tabular}
\caption{Preliminary analysis accuracy rates for media number 3}
\end{table}

\begin{table}
\begin{tabular}{|c|p{23.5em}|}
\hline
\multicolumn{2}{|l|}{Fully-Examined Case}\\ \hline
Suspect Objects & Notes\\ \hline
0 & No evidence or trace evidence relevant to the investigation\\ \hline
\end{tabular}
\caption{Results of a full examination on media number 4}
\end{table}

\begin{table}
\resizebox{\textwidth}{!} {
\begin{tabular}{|c|c|c|p{15em}|}
\hline
Examiner & Further Analysis & Suspect Objects & Notes\\ \hline
Examiner 5 & Yes & 45 & Decision made based on found images\\ \hline
Examiner 1 & No & 0 & n/a\\ \hline
Examiner 3 & No & 0 & n/a\\ \hline
Examiner 4 & No & 0 & Images found, but appear to be non-exploitation stock photos\\ \hline
Examiner 2 & No & 0 & n/a\\ \hline
\end{tabular}
}
\caption{Results of preliminary analysis on media number 4 from five examiners}
\end{table}

\begin{table}
\resizebox{\textwidth}{!} {
\begin{tabular}{|c|c|c|c|c|}
\hline
\multicolumn{5}{|l|}{Object Identification Error Rate}\\ \hline
Examiner False & Positive & False Positive Error & False Negative & False Negative Error\\ \hline
Examiner 5 & 45 & 1 & 0 & 0\\ \hline
Examiner 1 & 0 & 0 & 0 & 0\\ \hline
Examiner 3 & 0 & 0 & 0 & 0\\ \hline
Examiner 4 & 0 & 0 & 0 & 0\\ \hline
Examiner 2 & 0 & 0 & 0 & 0\\ \hline
\end{tabular}
}
\caption{Preliminary analysis object identification error rates for media number 4}
\end{table}

\begin{table}
\begin{tabular}{|c|c|c|c|}
\hline
\multicolumn{4}{|l|}{Accuracy Rate}\\ \hline
Examiner & Precision & Recall & F-measure\\ \hline
Examiner 5 & n/a & n/a & n/a\\ \hline
Examiner 1 & n/a & n/a & n/a\\ \hline
Examiner 3 & n/a & n/a & n/a\\ \hline
Examiner 4 & n/a & n/a & n/a\\ \hline
Examiner 2 & n/a & n/a & n/a\\ \hline
\end{tabular}
\caption{Preliminary analysis accuracy rates for media number 4}
\end{table}

\begin{table}
\begin{tabular}{|c|p{23.5em}|}
\hline
\multicolumn{2}{|l|}{Fully-Examined Case}\\ \hline
Suspect Objects & Notes\\ \hline
182 & More images appear to be one the machine but have yet to be categorized.\\ \hline
\end{tabular}
\caption{Results of a full examination on media number 5}
\end{table}

\begin{table}
\resizebox{\textwidth}{!} {
\begin{tabular}{|c|c|c|p{15em}|}
\hline
Examiner & Further Analysis & Suspect Objects & Notes\\ \hline
Examiner 4 & Yes & 66 & Decision made based on found images, keywords and encryption\\ \hline
Examiner 3 & Yes & 165 & Decision made based on found images, movies, keywords, Real Player history, evidence of disk wiping tools, evidence of encryption tools\\ \hline
Examiner 2 & Yes & 96 & Decision made based on found images, movies, encryption software, P2P, cleaner software\\ \hline
Examiner 5 & Yes & 31 & Decision made based on found images and movies\\ \hline
Examiner 1 & Yes & 85 & Decision made based on found images, movies\\ \hline
\end{tabular}
}
\caption{Results of preliminary analysis on media number 5 from five examiners}
\end{table}

\begin{table}
\resizebox{\textwidth}{!} {
\begin{tabular}{|c|c|c|c|c|}
\hline
\multicolumn{5}{|l|}{Object Identification Error Rate}\\ \hline
Examiner False & Positive & False Positive Error & False Negative & False Negative Error\\ \hline
Examiner 4 & 0 & 0 & 116 & .64\\ \hline
Examiner 3 & 0 & 0 & 16 & .09\\ \hline
Examiner 2 & 0 & 0 & 86 & .47\\ \hline
Examiner 5 & 0 & 0 & 151 & .83\\ \hline
Examiner 1 & 0 & 0 & 97 & .53\\ \hline
\end{tabular}
}
\caption{Preliminary analysis object identification error rates for media number 5}
\end{table}

\begin{table}
\begin{tabular}{|c|c|c|c|}
\hline
\multicolumn{4}{|l|}{Accuracy Rate}\\
Examiner & Precision & Recall & F-measure\\ \hline
Examiner 4 & 1 & .36 & .53\\ \hline
Examiner 3 & 1 & .91 & .95\\ \hline
Examiner 2 & 1 & .53 & .69\\ \hline
Examiner 5 & 1 & .17 & .29\\ \hline
Examiner 1 & 1 & .47 & .64\\ \hline
\end{tabular}
\caption{Preliminary analysis accuracy rates for media number 5}
\end{table}

\newpage

\end{document}